\documentstyle[12pt]{article}
\catcode`\@=11
\def\marginnote#1{}
\newcount\hour
\newcount\minute
\newtoks\amorpm
\hour=\time\divide\hour by60
\minute=\time{\multiply\hour by60 \global\advance\minute
by-\hour}\edef\standardtime{{\ifnum\hour<12
\global\amorpm={am}%
        \else\global\amorpm={pm}\advance\hour by-12 \fi
        \ifnum\hour=0 \hour=12 \fi
        \number\hour:\ifnum\minute<10
0\fi\number\minute\the\amorpm}}
\edef\militarytime{\number\hour:\ifnum\minute<10
0\fi\number\minute}

\def\draftlabel#1{{\@bsphack\if@filesw {\let\thepage\relax
   \xdef\@gtempa{\write\@auxout{\string
      \newlabel{#1}{{\@currentlabel}{\thepage}}}}}\@gtempa
   \if@nobreak \ifvmode\nobreak\fi\fi\fi\@esphack}
        \gdef\@eqnlabel{#1}}
\def\@eqnlabel{}
\def\@vacuum{}
\def\draftmarginnote#1{\marginpar{\raggedright\scriptsize\tt#1}}
\def\draft{\oddsidemargin -.5truein
        \def\@oddfoot{\sl preliminary draft \hfil
        \rm\thepage\hfil\sl\today\quad\militarytime}
        \let\@evenfoot\@oddfoot \overfullrule 3pt
        \let\label=\draftlabel
        \let\marginnote=\draftmarginnote

\def\@eqnnum{(\theequation)\rlap{\kern\marginparsep\tt\@eqnlabel}%
\global\let\@eqnlabel\@vacuum}  }
\def\numberbysection{\@addtoreset{equation}{section}
        \def\theequation{\thesection.\arabic{equation}}}

\def\underline#1{\relax\ifmmode\@@underline#1\else
 $\@@underline{\hbox{#1}}$\relax\fi}

\catcode`@=12
\relax

\numberbysection
\def\fin{\end{document}}
\def\beq{\begin{equation}}
\def\eeq{\end{equation}}
\def\bea{\begin{eqnarray}}
\def\eea{\end{eqnarray}}

\def\p{\Phi}

\def\Gcc#1 {{\cal G}^{[ #1 ]}_\|}
\def\Gcp#1 {{\cal G}^{[ #1 ]}_\bot}
\def\Cco#1 {{\cal C}^{[ #1 ]} }

\def\beq{\begin{equation}}
\def\eeq{\end{equation}}
\def\bea{\begin{eqnarray}}
\def\eea{\end{eqnarray}}
\def\p{\partial}
\def\l{\label}
\def\no{\nonumber}
\def\dh{\dot{h}}
\def\df{\dot{f}}
\begin{document}

\begin{titlepage}
\nopagebreak
\begin{flushright}
LPTENS-95/17\\
hep--th/9504131
\\
April 1995
\end{flushright}

\vglue 2.5  true cm
\begin{center}
{\large\bf
ON SOME INTEGRABLE GENERALISATIONS OF THE CONTINUOUS TODA SYSTEM}
\vglue 1  true cm
{\bf Mikhail V.  SAVELIEV}\footnote{
On leave of absence from the Institute for High Energy Physics,
142284, Protvino, Moscow region, Russia; saveliev@mx.ihep.su.
}\\
{\footnotesize Laboratoire de Physique Th\'eorique de
l'\'Ecole Normale Sup\'erieure\footnote{Unit\'e Propre du
Centre National de la Recherche Scientifique,
associ\'ee \`a l'\'Ecole Normale Sup\'erieure et \`a l'Universit\'e
de Paris-Sud.},\\
24 rue Lhomond, 75231 Paris C\'EDEX 05, ~France\footnote{
saveliev@physique.ens.fr}.}

\medskip
\end{center}

\vfill
\begin{abstract}
\baselineskip .4 true cm
\noindent
In the present paper we obtain some integrable generalisations of the
continuous Toda system generated by a flat connection
form taking values in higher grading subspaces of the algebra of the
area--preserving diffeomorphism of the torus $T^2$, and construct their
general solutions. The grading condition which we use here, imposed on
the connection, can be realised in terms of some holomorphic distributions
on the corresponding homogeneous spaces.
\end{abstract}
\vfill
\end{titlepage}
\baselineskip .5 true cm
\section{Introduction}
In the present paper we obtain some integrable generalisations
of the continuous Toda system generated by a flat connection form taking values
in higher grading subspaces of the Lie Poisson bracket algebra ${\cal L}$
on two--dimensional torus $T^2$. The main statement is the following.
Let ${\cal M}$ be a two--dimensional manifold with the local coordinates $z_+$
and $z_-$, and $G$ be an exponential
mapping of ${\cal L}$ defined, e.g., in the sense of inverse limit of
Hilbert.\footnote{Note that an exponential mapping of ${\cal L}$ gives a
differentiable structure which is weaker than that of a Lie group
in the classical sense. Nethertheless, in the same way as for $S_0\mbox{
Diff }T^2$, see e.g., \cite{ARS85}, it can be defined in the sense of this
limit, and just this fact allows us to perform a construction given
in the paper.} Then,

{\it A flat connection in the trivial fibre bundle ${\cal M}\times G
\rightarrow {\cal M}$ with $(1,0)$-- and $(0,1)$--components
taking values in the subspaces $\oplus_{m=0}^{m_+}{\cal G}_m$ and
$\oplus_{m=0}^{m_-}{\cal G}_{-m}$ of the ${\bf Z}$--graded Lie algebra
${\cal L}=\oplus_m{\cal G}_m$ generates an integrable system of
partial differential equations
\bea
& & \p_+\p_-\rho = \sum_{m=1}^{m_0}m\frac{\p^2}{\p\tau^2}
(\omega^+_m\omega^-_me^{m\rho}),\quad m_0\equiv \mbox{ min }(m_+,m_-);\l{eqs}\\
& & \p_{\mp}\omega^{\pm}_m  = \mp i\sum_{n=1}^{m_{\pm}-m}(m+n)
(\omega^{\pm}_{m+n})^{\frac{m}{m+n}}\frac{\p}{\p\tau}[
\omega^{\mp}_n(\omega^{\pm}_{m+n})^{\frac{n}{m+n}}e^{n\rho}], \no
\eea
$1\leq m\leq m_{\pm}-1$;
for the functions $\rho$ and $\omega^{\pm}_m$ depending on three
variables $z_{\pm}$ and $\tau$; $\omega^+_{m_+}=\omega^-_{m_-}\equiv 1$.
The general solution of the Goursat
(boundary value) problem to this sys\-tem is de\-ter\-mi\-ned by a com\-ple\-te
set of arbi\-tra\-ry fun\-cti\-ons $\Phi^+_m(z_+; \tau ), \, 1\leq m\leq m_+$,
and $\Phi^-_m(z_-; \tau ), \, 1\leq m\leq m_-$.}

In differential geometry language, the grading condition we use, imposed
on the flat connection, can be realised, modifying the reasonings given
in \cite{RS94} for finite--dimensional case, in terms of some holomorphic
and antiholomorphic distributions on the homogeneous spaces
 $F_+=G/B_-$ and $F_-=G/B_+$, respectively, where
$B_{\pm}$ are exponential mappings of $\oplus_{m=0}^{\infty}{\cal G}_{\pm
m}$. Such a distribution, say on $F_+$, is generated by the subspaces of
$T^{(1,0)}(F_+)$ with the help of the canonical projection $\pi_+:G
\rightarrow F_+$, as $\pi_{+*}({\cal M}_{+h})$. Here ${\cal M}_{+h}$ is
obtained by the left translation (with an element $h$ of $G$, $\pi_+(h)\in
F_+$) of the subspace ${\cal M}_+\equiv \oplus_{m\leq m_+}{\cal G}_{+m}$
invariant under the adjoint action of $B_-$ in ${\cal L}$. A distribution
on $F_-$ is defined similarly, in terms of the canonical projection $\pi_-:G
\rightarrow F_-$, and the subspace ${\cal M}_-\equiv \oplus_{m\leq m_-}{\cal
G}_{-m}$.

For the simplest case when the corresponding connection takes values
in the local part ${\cal G}_{-1}\oplus {\cal G}_0\oplus {\cal G}_{+1}$
of ${\cal L}$, equations (\ref{eqs}) are reduced to the
well--known continuous Toda system (the heavenly equation familiar thanks
to \cite{BF82})
\beq
\p_+\p_-\rho = \frac{\p^2}{\p\tau^2}e^{\rho};\l{heavenly}
\eeq
for which we briefly discuss the boundary condition on a half-line $z_+=z_-$.
Quite significative, while still rather simply represented case is with
$m_{\pm}=2$, when system (\ref{eqs}) is reduced to the following equations:
\bea
& & \p_+\p_-\rho = \frac{\p^2}{\p\tau^2}[2e^{2\rho}+
\omega^+\omega^-e^{\rho}],\no \\
& & \p_{\mp}\omega^{\pm}  =  \mp 2i\frac{\p}{\p\tau}[\omega^{\mp
}e^{\rho}] \l{eqs2}.
\eea

As a direct by--product of the given consideration one obtains the integrable
equations corresponding to the higher grading subspaces of the centreless
Virasoro--Witt algebra.

Note that the prob\-lem of con\-struc\-ting non\-li\-near inte\-gra\-ble
two--\-di\-men\-si\-onal par\-tial dif\-fe\-ren\-tial equa\-ti\-ons associated
with higher ${\bf Z}$--grading subspaces has been already considered in a
general and abstract form, in fact, for an arbitrary Lie algebra ${\cal G}$,
see \cite{LS92}. However, even in the simplest case like $A_n$, the
arising equations and their general solutions look, in a sense, more
complicated than their continuous limit (\ref{eqs}).

\section{Some In\-for\-ma\-tion About Lie Pois\-son Bra\-cket Algebra}

Let us recall briefly some information about Lie Poisson bracket
algebra ${\cal L}$ following \cite{SV89}, \cite{S92}. This algebra,
being considered as ${\bf Z}$--graded continuum Lie algebra ${\cal G}(E;
-i\p/\p \tau,-i\p/\p \tau )=\mathop{\oplus}\limits_{m \in {\bf Z}}{\cal G}_m$,
is isomorphic to $A_{\infty}$ and $S_0\mbox{ Diff }T^2$ (infinitesimal
area--preserving diffeomorphisms of the torus $T^2$). Its elements satisfy
the commutation relations
\beq
{}[X_m(\phi ), X_n(\psi )]=iX_{m+n}(n\dot{\phi}\psi -m\phi\dot{\psi}).
\l{cr}
\eeq
Here $X_m(\phi)=\int d\tau X_m (\tau )\,\phi (\tau )$ are the elements of the
subspaces ${\cal G}_m$ pa\-ra\-met\-rised by the functions $\phi(\tau )$
belonging to the algebra $E$ of trigonometrical polynomials on a circle;
$\dot{\phi} \equiv \frac{\p\phi}{\p\tau}$.

The algebra ${\cal G}(E;-i \frac{\p}{\p\tau};-i \frac{\p}{\p\tau})$ is of
constant growth (in a functional sense) since ${\cal G}_n \simeq {\cal G}_1
\simeq E$; its
Cartan subalgebra $\wp \simeq {\cal G}_0$ is infinite--dimensional; the roots
are $n\delta'(\tau)$. Let $\wp^*$ be an algebra dual to $\wp$, let $V$ be a
${\cal L}$--module and $\lambda \in \wp^*$. Denote by $V_\lambda$ a set of
vectors $v \in V$ satisfying $X_0(\phi )v = \lambda (\phi )v$ for all $\phi
\in E$. It can be shown by an appropriate limit procedure
(starting from $A_r$ and using the aforementioned isomorphism ${\cal G}(E;-i
\frac{\p}{\p\tau};-i \frac{\p}{\p\tau}) \simeq {\cal
G}(E;-\frac{\p ^2}{\p\tau^2};
\mbox{ id }) \simeq A_\infty$) that there exists a nonzero vector ${\bar v}\in
V$ such that ${\cal G}_{m}(\bar v) = 0$ for $m>0$ and $U({\cal L})(\bar v)= V$.
Here $U(\cal L)$ is the universal enveloping algebra for ${\cal
L}$. By analogy with the usual (``discrete'') case, this ${\cal L}$--module
$V$ is called the highest weight module, and $\bar v$ the highest weight
vector.

A symmetrical bilinear invariant form on the local part of the algebra in
question is defined as follows
\[
\mbox{ tr }(X_i(f)X_j(g)) = \delta_{i+j,0}(f, g), \qquad i,j=0, \pm 1; \]
where
\[
(f, g)\equiv\int d\tau f(\tau )g(\tau ).\]
Thanks to the invariance property of the form,
\[
([X_m(f), X_{-m}(g)], X_0(h))=(X_m(f), [X_{-m}(g), X_0(h)]),\]
the commutation relations (\ref{cr}) give
\[
-im(X_0(\frac{\p}{\p\tau}(fg), X_0(h))=im(X_m(f), X_{-m}(g\dh ));\]
and so one has
\[
(X_m(f), X_{-m}(g\dh))=-\int d\tau\frac{\p}{\p\tau}(f(\tau )
g(\tau ))h(\tau )=\int d\tau fg\dh .\]
Thereof,
\beq
(X_m(f), X_n(g))=\delta_{m+n,0} (f, g).\l{sp}
\eeq
To this end denote a continuous version (in the sense of the
algebra $A_{\infty}$ with the elements $X^{(A)}_m$) of the
highest weight vectors of the fundamental representations of $A_n$ by $\vert
\tau >$, for which
\beq
X^{(A)}_0(\phi )\,\vert \tau >=\phi (\tau )\vert \tau >, \;
X^{(A)}_m(\phi )\,\vert \tau >=0 \mbox{ for } m\geq 1;
\l{frA}
\eeq
so that for the case of ${\cal L}$ one has
\beq
X_0(\phi )\,\vert \tau >=\p_{\tau}^{-1}\phi (\tau )\vert \tau >, \;
X_m(\phi )\,\vert \tau >=0 \mbox{ for } m\geq 1.
\l{fr}
\eeq

\section{Derivation of the Equations}

Let ${\cal M}$ be a two--dimensional manifold with the
local coordinates $z_+$ and
$z_-$ in ${\cal M}$, and ${\cal A}$ be a Lie algebra ${\cal G}$--valued
$1$--form on ${\cal M}$, ${\cal A}={\cal A}_+dz_++{\cal A}_-dz_-$. Any such
$1$--form generates a connection form of some connection in the trivial fibre
bundle. Suppose that ${\cal A}$ is flat, so that ${\cal A}_{\pm}$ are some
mappings from ${\cal M}$ to ${\cal G}$, satisfying the zero curvature condition
\beq
{}[\frac{\p}{\p z_+}+{\cal A}_+, \frac{\p}{\p z_-}+{\cal A}_-]=0.  \l{zc}
\eeq
Let the connection ${\cal A}$ takes values in the algebra ${\cal L}$.
Choosing the basis $X_m(\phi )$ in ${\cal G}(E;-i\frac{\p}{\p\tau};-i
\frac{\p}{\p\tau})$ and considering the components of the decomposition
of ${\cal A}_{\pm}$ over this basis as fields, we can treat (\ref{zc}) as
a nonlinear system of partial differential equations for the fields.
To provide nontriviality of such a system, we impose the grading
condition on the connection \cite{LS92}, such that the $(1,0)$--component
${\cal A}_+$ of ${\cal A}$ takes values in $\oplus _{m\geq 0}{\cal G}_{+m}$,
and the $(0,1)$--component ${\cal A}_-$ takes values in $\oplus _{m\geq 0}
{\cal G}_{-m}$. Namely,
\beq
{\cal A}_+=\sum_{m=0}^{m_+}X_m(f^+_m),\qquad
{\cal A}_-=\sum_{m=0}^{m_-}X_{-m}(f^-_m). \l{spec}
\eeq
Then substituting (\ref{spec}) in (\ref{zc}), one has
\bea
& & X_0\{ \p_+f^-_0-\p_-f^+_0-i\sum_{m=1}^{m_0}m(f^+_mf^-_m)^{.}\} =0,
\l{oe}\\
& & \sum_{m=1}^{m_+}X_m\{ \p_-f^+_m+imf^+_m\df^-_0+i\sum_{1\leq
n<m\leq m_+}X_{m-n}(n\df^+_mf^-_n+mf^+_m\df^-_n)\}=0,\no \\
& & \sum_{m=1}^{m_-}X_{-m}\{ \p_+f^-_m-imf^-_m\df^+_0-i\sum_{1\leq
n<m\leq m_-}X_{-m+n}(n\df^-_mf^+_n+mf^-_m\df^+_n)\}=0;\no
\eea
or, in terms of the functions $f^{\pm}(z_+,z_-)$,
\bea
& & \p_+f^-_0-\p_-f^+_0=i\sum_{m=1}^{m_0}m\frac{\p}{\p\tau}(f^+_mf^-_m)
=0;\l{E}\\
& & \p_-f^+_m+imf^+_m\df^-_0=-i\sum_{n=1}^{m_+-m}[n\df^+_{m+n}f^-_n
+(m+n)f^+_{m+n}\df^-_n],\quad 1\leq m\leq m_+;\no \\
& & \p_+f^-_m-imf^-_m\df^+_0=i\sum_{n=1}^{m_--m}[n\df^-_{m+n}f^+_n
+(m+n)f^-_{m+n}\df^+_n].\quad 1\leq m\leq m_-.\no
\eea
It follows from the last two systems with $m=m_+$ and
$m=m_-$, respectively, that
\beq
\df^-_0=\frac{i}{m_+}\p_-\mbox{ ln }f^+_{m_+},\quad
\df^+_0=-\frac{i}{m_-}\p_+\mbox{ ln }f^-_{m_-};\l{f0}
\eeq
and from the other equations of these systems
\bea
& & f^+_m\p_-\mbox{ ln }[f^+_m(f^+_{m_+})^{-\frac{m}{m_+}}]=\no \\
& &-i\sum_{n=1}^{m_+-m}f^+_{m+n}f^-_n \frac{\p}{\p\tau}\mbox{ ln
}[(f^+_{m+n})^n(f^-_n)^{m+n}],\; 1\leq m\leq m_+-1;\l{f+} \\
& & f^-_m\p_+\mbox{ ln }[f^-_m(f^-_{m_-})^{-\frac{m}{m_-}}]=\no \\
& & i\sum_{n=1}^{m_--m}f^-_{m+n}f^+_n\frac{\p}{\p\tau} \mbox{ ln
}[(f^-_{m+n})^n(f^+_n)^{m+n}],\; 1\leq m\leq m_--1.\l{f-}
\eea
Substituting (\ref{f0}) in the first equation (\ref{E})
differentiated over $\tau$, one has
\beq
\p_+\p_-\mbox{ ln }[(f^+_{m_+})^{1/m_+} (f^-_{m_-})^{1/m_-}]=
\sum_{m=1}^{m_0}m\frac{\p^2}{\p\tau^2}f^+_mf^-_m.\l{00}
\eeq
Then, with the notations
\beq
f^{\pm}_m(f^{\pm}_{m_{\pm}})^{-\frac{m}{m_{\pm}}}\equiv \omega^{\pm}_m,
\; 1\leq m\leq m_{\pm}-1;\; (f^+_{m_+})^{1/m_+} (f^-_{m_-})^{1/m_-}
\equiv e^{\rho},\l{notat}
\eeq
our equations are finally written as (\ref{eqs}). In the simplest case when
$m_{\pm}=1$, this system is reduced to equation (\ref{heavenly}), whose
general solution is determined \cite{S89} by two arbitrary functions
$\Phi^+(z_+,\tau )$ and $\Phi^-(z_-,\tau )$; for $m_{\pm}=2$ -- to equations
(\ref{eqs2}).

The corresponding connection components in terms of the
functions $\rho$ and $\omega^{\pm}_m$ are rewritten as
\bea
{\cal A}_{\pm} & = & X_0(\mp\frac{i}{m_{\mp}}\p_{\pm}\p^{-1}_{\tau}\mbox{ ln
}f^{\mp}_{m_{\mp}})\no \\
& + & \sum_{m=1}^{m_{\pm}-1}X_{\pm m}(\omega^{\pm}_m
(f^{\pm}_{m_{\pm}})^{\frac{m}{m_{\pm}}})+
X_{\pm m_{\pm }}(f^{\pm}_{m_{\pm}}); \l{A}
\eea
or, in an appropriate gauge, as
\beq
{\cal A}_{\pm}=X_0(\mp\frac{i}{2}\p_{\pm}\p^{-1}_{\tau}\rho )+
\sum_{m=1}^{m_{\pm}-1}X_{\pm m}(\omega^{\pm}_me^{\frac{m}{2}\rho})+
X_{\pm m_{\pm }}(e^{\frac{m_{\pm}}{2}\rho}).\l{Ag}
\eeq

For the case when all the functions depend on $\tau$ and $z_+-z_-$ only,
putting $-i\p_++i\p_-=\frac{1}{2}\frac{\p}{\p r}$ and choosing a
gauge with $f^{\pm}_{m_{\pm}}=e^{\rho}$, for e.g. $m_{\pm}=2$
we have the Lax operator
\bea
L & \equiv & {\cal A}_++{\cal A}_-=X_0(\p_r\p^{-1}_{\tau}\rho ) \l{L} \\
& + & X_1(\omega^+e^{\rho /2})
+X_{-1}(\omega^-e^{\rho /2})+X_2(e^{\rho})+X_{-2}(e^{\rho}).\no
\eea

Then one can calculate, using (\ref{sp}), all conserved quantities as
${\cal I}_p\equiv
\frac{1}{p}\mbox{ tr }L^p$, for which $\frac{\p}{\p r}{\cal I}_p=0$, e.g.
\beq
{\cal I}_2=\int d\tau [\frac{1}{2}(\frac{\p y}{\p r})^2+e^{\frac{\p y}{\p
\tau}}(e^{\frac{\p y}{\p \tau}}+\omega^+\omega^-)],\l{I2}
\eeq
where $y=\p^{-1}_{\tau}\rho$.

Finishing the section, let us note that equations (\ref{eqs}), remaining
integrable, admit the natural reduction to a system associated with the higher
grading subspaces of the centreless Virasoro--Witt algebra ${\cal W}$. Indeed,
it is clear that with a choice $\phi (\tau)\simeq\tau$, the elements $X_m
(\phi )$ of the Lie Poisson bracket algebra ${\cal L}$ given by the commutation
relations (\ref{cr}), generate the algebra ${\cal W}$ with the elements $L_m$
satisfying the relations
\beq
{} [L_m, L_n]=(m-n)L_{m+n}.\l{vw}
\eeq
Then equations (\ref{eqs}) take the form
\bea
& & \p_+\p_-\rho = 2\sum_{m=1}^{m_0}m
\omega^+_m\omega^-_me^{m\rho};\l{vir}\\
& & \p_{\mp}\omega^{\pm}_m  = \mp i\sum_{n=1}^{m_{\pm}-m}(m+2n)
\omega^{\mp}_n \omega^{\pm}_{m+n}e^{n\rho},\quad
1\leq m\leq m_{\pm}-1;\no
\eea
which for the special case $m_+=m_-=2$ coincides with those in \cite{LLS82}.

\section{Construction of the General Solution}

Let us construct the general solution to system (\ref{eqs}) using an
appropriate modification of the method given in \cite{LS92}.
The connection ${\cal A}$ is represented in the gradient
form,
\beq
{\cal A}_{\pm}=g^{-1}\p_{\pm}g,\l{cm}
\eeq
with $g\in G$; and we take for ${\cal A}_{\pm}$
the Gauss type decomposition of $g$,\footnote{It seems
believable that such a decomposition for $G$ can be considered as an
appropriate continuous limit of the Gauss decomposition for $A_r$.}
\[
g=M_-N_+g_{0-} \mbox{ and } g=M_+N_-g_{0+},\]
respectively. The grading conditions
(\ref{spec}) provide the holomorphic property of $M_{\pm}$, namely
that  the functions
$M_{\pm}(z_{\pm})\in G_{\pm}$ satisfy the initial value problem
\beq
\p_{\pm}M_{\pm}(z_{\pm})=M_{\pm}(z_{\pm})L_{\pm}(z_{\pm}),\l{ivp}
\eeq
where
\beq
L_{\pm}(z_{\pm})=\sum_{m=1}^{m_{\pm}}X_{\pm m}(\Phi^{\pm}_m(z_{\pm}))
\l{arb}
\eeq
with arbitrary functions $\Phi^{\pm}_m(z_{\pm};\tau )$ determining the
general solution to our system.  Then one has
\beq
{\cal A}_{\pm}=g_{0\mp}^{-1}(N_{\pm}^{-1}\p_{\pm}N_{\pm})g_{0\mp}+
g_{0\mp}^{-1}\p_{\pm}g_{0\mp}.
\l{A+-}
\eeq
Differentiating the identity
\beq
M_+^{-1}M_-=N_-g_0N_+^{-1}, \l{id}
\eeq
with $g_0\equiv g_{0+}g_{0-}^{-1}$,
over $z_{\pm}$  and using equations (\ref{ivp}), one can  get convinced
that the elements $N_{\pm}^{-1}\p_{\pm}N_{\pm}$ take values in the subspaces
$\oplus_{m=1}^{m_{\pm}}{\cal G}_{\pm m}$. Rewrite now decomposition
(\ref{A+-}) as
\beq
{\cal A}_{\pm}=g_{0\pm}^{-1}\tilde{L}_{\pm}g_{0\pm}+
g_{0\mp}^{-1}\p_{\pm}g_{0\mp},\l{AG}
\eeq
where
\[
\tilde{L}_{\pm} \equiv g_{0}^{\pm 1}(N_{\pm}^{-1}\p_{\pm}N_{\pm})g_{0}^{\mp 1}
 =\sum_{m=1}^{m_{\pm}}X_{\pm m}(F^{\pm}_m)\]
with some functions $F^{\pm}_m(z_+, z_-)$. In the same way as for the case
of a simple Lie algebra, here
the functions $\tilde{L}_{\pm}$ are equal to $L_{\pm}$ only
when $m_+=m_-=1$.
However, thanks to identity (\ref{id}), the elements $N_{\pm}$ and $g_0$
are determined by the elements $M_{\pm}$, and hence the functions $F^{\pm}_m$
entering
$\tilde{L}_{\pm}$ can be expressed in terms of the functions
$\Phi^{\pm}_m(z_{\pm})$ for arbitrary values of $m_{\pm}$; moreover,
it is clear that
$F^{\pm}_{m_{\pm}}=\Phi^{\pm}_{m_{\pm}}$.  A construction of the general
solution
to system (\ref{eqs}) can be done in an explicit way. Before we give it,
let us discuss briefly the solution to equations (\ref{ivp}).

The functions $M_{\pm}(z_{\pm})\in G_{\pm}$ satisfying the initial value
problem (\ref{ivp}) are given by the corresponding multiplicative integrals
\beq
M_{\pm}=\sum_{n=0}^{\infty}\int_{Z^n_{\pm}(z_{\pm})}dy_{\pm}
L_{\pm}(y_{\pm}^{n})
L_{\pm}(y_{\pm}^{n-1})\cdots L_{\pm}(y_{\pm}^{1}),\l{mi1}
\eeq
or, equivalently,
\beq
M_{\pm}^{-1}=\sum_{n=0}^{\infty}(-1)^n\int_{Z^n_{\pm}(z_{\pm})}dy_{\pm}
L_{\pm}(y_{\pm}^{1})L_{\pm}(y_{\pm}^{2})\cdots L_{\pm}(y_{\pm}^{n}),\l{mi2}
\eeq
where $Z^n_{\pm}(z_{\pm})=\{ y_{\pm}\in {\bf R}^n: a_{\pm}\leq y_{\pm}^{n}
\leq y_{\pm}^{n-1}\leq \cdots \leq y_{\pm}^{1}\leq z_{\pm}\}$; $a_{\pm}$ are
some constants determining the problem; $M_{\pm}(a_{\pm})=1$. Note that
there is also a noncommutative version of the well--known exponential formula,
see  \cite{Str87}, \cite{LS92},
\bea
M_{\pm} & = & \exp
\sum_{n=1}^{\infty}\sum_{\omega}\frac{(-1)^{\epsilon (\omega )}}{n^2
C^{n-1}_{\epsilon (\omega )}}\times\no \\
& & \int_{Z^n_{\pm}}dy_{\pm}[\cdots [L_{\pm}(y_{\pm}^{\omega (n)}),
L_{\pm}(y_{\pm}^{\omega (n-1)})]\cdots ]L_{\pm}(y_{\pm}^{\omega (1)}].
\l{bcdh}
\eea
Here $\omega$ is a permutation of the set $\{ 1, 2, \cdots ,
n\}$; $\epsilon (\omega )$ is the number of errors in the ordering
consecutive terms in $\{ \omega (1), \omega (2), \cdots , \omega (n)\}$;
$C^{n-1}_{\epsilon (\omega )}$ are the binomial coefficients.

The algebra ${\cal G}(E;-i\p/\p \tau,-i\p/\p \tau )$, which we have introduced
as a ${\bf Z}$--graded algebra, can be re\-pre\-sen\-ted as ${\bf
Z}^2$--gra\-ded alge\-bra with one--di\-men\-si\-onal components, e.g., with
the elements $X_{\bf m}$ satisfying the commutation relations \cite{A66}
\beq
{}[X_{\bf m}, X_{\bf n}]=({\bf m}\times {\bf n})X_{{\bf m}+{\bf n}},
\l{dis}
\eeq
where ${\bf m}=(m_1,m_2)$ and ${\bf n}=(n_1,n_2)$ are two--di\-men\-si\-onal
integer vectors; $({\bf m}\times {\bf n})\equiv m_1n_2-m_2n_1$. With account
of this fact, the multiple commutators of the functions
\[
L(y^{(k)})\equiv\sum_{m_1^{(k)}=1}^{m_{\pm}}X_{m_1^{(k)}}(\Phi
^{(k)}_{m_1^{(k)}})\]
in the exponential (\ref{bcdh}) are given by the formula
\bea
& & [\cdots [L(y^{(n)}), L(y^{(n-1)})]\cdots ]L(y^{(1)}]=\no\\
& & \sum_{m_2^{(1)},\cdots ,m_2^{(n)}}\Phi_{{\bf m}^{(n)}}^{(n)}
\prod_{l=1}^{n-1}((\sum_{j=l+1}^n{\bf m}^{(j)})\times {\bf m}^{(l)})
\Phi_{{\bf m}^{(l)}}^{(l)}
X_{\sum_{j=1}^n{\bf m}^{(j)}},\l{sum}
\eea
where
\[
\Phi^{(k)}_{m_1^{(k)}}(z;\tau )=\sum_{m_2^{(k)}}\Phi^{(k)}_{{\bf
m}^{(k)}}(z)e^{im_2^{(k)}\tau}.\]

Now let us give an explicit solution to our system (\ref{eqs}).
Equating expressions (\ref{Ag}) and (\ref{AG}), we have
\beq
g_{0\mp}^{-1}\p_{\pm}g_{0\mp}=X_{0}(\mp\frac{i}{2}\p_{\pm}
\p_{\tau}^{-1}\rho ),\l{g0}
\eeq
thereof
\[
g_{0\mp}=g^{\mp}_0(z_{\mp})e^{\mp\frac{i}{2}X_0(\p_{\tau}^{-1}\rho
)},\]
and hence
\[
g_0=g^+_0(z_+)e^{iX_0(\p_{\tau}^{-1}\rho )}(g^-_0(z_-))^{-1}.\]
Here $g^{\pm}_0\equiv g^{\pm}_0(z_{\pm})\in G_0$ are arbitrary
functions of their arguments, and are expressed
in terms of $\Phi^{\pm}_m$.  Moreover,
\bea
g_{0\pm}^{-1}\tilde{L}_{\pm}g_{0\pm}& = & e^{\mp\frac{i}{2}X_0(\p_{\tau}^{-1}
\rho )}
(g^{\pm}_0)^{-1}\sum_{m=1}^{m_{\pm}}X_{\pm m}(F^{\pm}_m)g^{\pm}_0
e^{\pm\frac{i}{2}X_0(\p_{\tau}^{-1}\rho )}\no\\
& = & \sum_{m=1}^{m_{\pm}-1}X_{\pm m}(\omega^{\pm}_me^{\frac{m}{2}\rho})+
X_{\pm m_{\pm }}(e^{\frac{m_{\pm}}{2}\rho}).\l{g+}
\eea

Rewriting identity (\ref{id}) in the form
\[
(g^+_0)^{-1}M_+^{-1}M_-g^-_0=[(g^+_0)^{-1}N_-g^+_0]
e^{iX_0(\p_{\tau}^{-1}\rho )}[(g^-_0)^{-1}N_+^{-1}g^-_0],\]
one has
\beq
<\tau \vert (g^+_0)^{-1}M_+^{-1}M_-g^-_0\vert\tau >=<\tau\vert
e^{iX_0(\p_{\tau}^{-1}\rho )}\vert\tau >, \l{hv}
\eeq
where the brackets are taken between the basis vector $\vert\tau >$
and its dual, $< \tau\vert$,
annihilated by the subspaces ${\cal G}_+$ and ${\cal G}_-$, respectively.
This matrix element realises a continuous
version of the tau--function depending on the necessary number of arbitrary
functions $\Phi^{\pm}_m(z_{\pm},\tau )$ which determine the general solution
to system (\ref{eqs}). It can be rewritten as
a series over the nested integrals of the products of these
functions in the same way as it was done for the case of the continuous
Toda system (\ref{heavenly}) in terms of the basis (\ref{fr})
with the help of the commutation relations (\ref{cr}), see e.g., \cite{S92},
or with the discrete basis (\ref{dis}) and formulae (\ref{sum}).

Now, since, in accordance with (\ref{cr}),
\beq
e^{X_0(\Phi )}X_m(\Psi )e^{-X_0(\Phi )}=X_m(e^{im\p_{\tau}\Phi}\Psi ),
\l{rot}
\eeq
one has from (\ref{g+}) that
\[
(g^{\pm}_0)^{-1}\sum_{m=1}^{m_{\pm}}X_{\pm m}(F^{\pm}_me^{\frac{m}{2}
\rho })g^{\pm}_0=\sum_{m=1}^{m_{\pm}-1}X_{\pm m}(\omega^{\pm}_me^{\frac{m}{2}
\rho })+X_{\pm m_{\pm}}(e^{\frac{m_{\pm}}{2}\rho});\]
and hence
\beq
\sum_{m=1}^{m_{\pm}}X_{\pm m}(F^{\pm}_m)=\sum_{m=1}^{m_{\pm}-1}
g_0^{\pm}X_{\pm m}(\omega^{\pm}_m)(g_0^{\pm})^{-1}+
g_0^{\pm}X_{\pm m_{\pm}}(g_0^{\pm})^{-1}.\l{rel}
\eeq
Thanks to (\ref{rel}), with the parametrisation
$g_0^{\pm}=e^{X_0(\p_{\tau}^{-1}\nu_{\pm})}$, one has
$F^{\pm}_{m_{\pm}}=\Phi^{\pm}_{m_{\pm}}=e^{\pm im_{\pm}\nu_{\pm}}$,
so that
\beq
g_0^{\pm}(z_{\pm})=e^{\mp\frac{i}{m_{\pm}}X_0(\p_{\tau}^{-1}\mbox{
ln }\Phi^{\pm}_{m_{\pm}}(z_{\pm}))}.\l{p0}
\eeq
Now expression (\ref{hv}) implies, with account of (\ref{fr}) and
(\ref{p0}), that
\beq
e^{\rho}=(\Phi^+_{m_+})^{\frac{1}{m_+}}(\Phi^-_{m_-})^{\frac{1}{m_-}}
e^{-i\p_{\tau}^2\mbox{ ln }<\tau\vert M_+^{-1}M_-\vert\tau >},\l{me}
\eeq
where $M_{\pm}(z_{\pm})$ are determined from the initial value problem
(\ref{ivp}) by the infinite series (\ref{mi1}), (\ref{mi2}), or
exponentials (\ref{bcdh}). Here the continuous analogue of the tau--function
$<\tau\vert M_+^{-1}M_-\vert\tau >$ is represented by the series
\bea
& & \sum_{m,n=0}^{\infty}(-1)^n\int_{Z^n_{+}(z_{+})}dy_+\int_{Z^m_{-}(z_{-})}
dy_-\times \no \\
& & <\tau\vert L_+(y_+^1)\cdots L_+(y_+^n)L_-(y_-^m)L_-(y_-^1)
\vert\tau >,\l{prel}
\eea
where nonzero contributions come only from the terms with the
same number of the elements $L_+$ and $L_-$ in their products in
(\ref{prel}), i.e.,
\bea
<\tau\vert M_+^{-1}M_-\vert\tau > & & = 1+\sum_{n\geq 1}
(-1)^n\int_{Z^n_{+}(z_{+})}dy_+\int_{Z^n_{-}(z_{-})}
dy_-\times \no \\
& & <\tau\vert L_+(y_+^1)\cdots L_+(y_+^n)L_-(y_-^n)L_-(y_-^1)
\vert\tau >.\l{tau}
\eea
The rule for explicit calculating the matrix elements in series (\ref{tau})
consists in successive displacements of all $L_+$ and $L_-$ to
the extreme right and, respectively, left position at which they
annihilate the highest vectors $\vert\tau >$ and $<\tau\vert$.
In this one uses the decomposition of $L_{\pm}$ as series (\ref{arb}),
and relations (\ref{fr}) and (\ref{cr}).

Moreover, from (\ref{rel}) we obtain
\[
X_{\pm m}(F^{\pm}_m)=g_0^{\pm}X_{\pm m}(\omega^{\pm}_m)(g_0^{\pm})^{-1}=
X_{\pm m}(\omega^{\pm}_me^{\frac{m}{m_{\pm}}\mbox{ ln }\Phi^{\pm}_{m_{\pm}}}),
\]
$1\leq m\leq m_{\pm}-1$, and hence
\beq
\omega^{\pm}_m=F^{\pm}_m(\Phi^{\pm}_{m_{\pm}})^{-\frac{m}{m_{\pm}}},
\quad 1\leq m\leq m_{\pm}-1.\l{fin}
\eeq
So, formulas (\ref{me}) and (\ref{fin}) define the general
solution to system (\ref{eqs}).

The functions $F^{\pm}_m, \,   1\leq m\leq m_{\pm}-1$, entering solution
(\ref{fin}), and
in turn $\tilde{L}_{\pm}$, are determined in terms of the matrix elements of
the known
element $M_+^{-1}M_-$ taken between some, not necessarily highest
vectors of the
representation space. Consider, for example, the matrix element
with the highest bra-vector
$<\tau\vert$ and a ket-vector $\vert\tau >^{(1)}$ which is
annihilated by the action of the
subspaces ${\cal G}_m,\, m>1$. Then
\[
<\tau\vert g_0^{-1}M_+^{-1}M_-\vert\tau >^{(1)}= <\tau\vert
(g_0^{-1}N_-g_0)N_+^{-1}\vert\tau
>^{(1)}=<\tau\vert N_+^{-1}\vert\tau >^{(1)},\]
and differentiating this equality over $z_+$, one has a sequence of equalities
\bea
\p_+<\tau\vert g_0^{-1}M_+^{-1}M_-\vert\tau >^{(1)} & = &
-<\tau\vert  g_0^{-1}\tilde{L}_+g_0
N_+^{-1}\vert\tau >^{(1)}=\nonumber\\
-<\tau\vert  g_0^{-1}X_1(F^+_1)g_0N_+^{-1}\vert\tau >^{(1)} & = &
-<\tau\vert  g_0^{-1}X_1(F^+_1)g_0\vert\tau >^{(1)},\nonumber
\eea
which determines the function $F^+_1$. Now, knowing this function, to
find $F^+_2$ we consider
the matrix element with  a ket-vector $\vert\tau >^{(2)}$ which is
annihilated by the action of the
subspaces ${\cal G}_m,\, m>2$; etc. The similar procedure allows to
determine the functions
$F^-_m$.

It seems interesting to investigate the problem of the integrability of system
(\ref{eqs}) on a half--line $r=0$. Let us study it by the simplest example of
the heavenly equation (\ref{heavenly}) which shows that this problem
respects the boundary condition
\beq
\p_r x=0 \quad \mbox{at } r=0; \quad \rho \equiv \p^2_{\tau} x. \l{bc0}
\eeq
To get convinced in it one can equate the corresponding
$W^{\pm}_m$-elements,
$\p_{\mp}W^{\pm}_m=0$, at $z_+=z_-$. Then it follows from the expression
\[
W^{\pm}_2=\int d\tau [\p^2_{\pm}x+\frac{1}{2}(\p_{\pm}\p_{\tau}x)^2],\]
that
\beq
\p_r x=\lambda\cdot e^{\rho /2} \quad \mbox{at } r=0, \l{bc}
\eeq
where $\lambda$ is an arbitrary constant yet; formally the similar condition
takes place for the case of the Liouville equation \cite{GN82}. Then, already
from the $W^{\pm}$-elements of the 3rd order,
\[
W^{\pm}_3=\int d\tau [\tau\p^3_{\pm}x+(\p_{\pm}x)\cdot (\p_{\pm}^2\p_{\tau}x)+
\tau (\p_{\pm}\p_{\tau}x)\cdot (\p_{\pm}^2\p_{\tau}x)
-\frac{1}{3}(\p_{\pm}\p_{\tau}x)^3],\]
one sees that this constant $\lambda$ should be equal to zero. Recall  that
for the affine Toda field theory associated with the series $A_n^{(1)}$,
as well as it can be easily shown for the corresponding finite
case $A_n$, there are two possibilities for constants $\lambda_i$ entering the
boundary conditions like (\ref{bc}), or its affine deformation, written for the
fields $x_i, \; 1\leq i\leq r$; namely $\vert\lambda_i\vert =2$, or all
$\lambda_i =0$ \cite{CDRS94}.\footnote{The same as for the Liouville case, the
sine (or sinh)--Gordon model is less restricted in this sense; see
\cite{CDRS94} for the relevant references.}  It is
interesting to note that nonabelian Toda systems, in general, admit not such a
rigid restriction on these constants. In particular, for the case of $B_2$,
the boundary conditions contain one arbitrary constant, as it can be seen
by equating the corresponding $W^{\pm}$-elements, all three of them being
of the $2$nd order \cite{S90}.

\section{Algebraic Structure of the Solution}

The matrix element (\ref{tau}) determining the general solution to system
(\ref{eqs}) realises a continuous version of the well--known object in the
theory of integrable systems -- the tau--function. At the same time this
matrix element is closely related to the so--called Shapovalov form defined
here on the Lie algebra ${\cal L}$, that can clarify an algebraic structure
of the solution. Such a relationship is quite common and takes place
for a wide class of nonlinear integrable systems, including, in particular,
abelian and nonabelian Toda systems associated with the simple
Lie algebras ${\cal G}$; let us discuss it in brief.

Recall, see e.g. \cite{Zh94}, that the standard Shapovalov form defines the
linear mapping $U({\cal G})\otimes_{\bf C} U({\cal G})\longmapsto U({\wp })$
and is realised as a bilinear form $(x^{\vee} y)_0$ for any two elements $x,
y\in U({\cal G})$. Here $x\rightarrow x^{\vee}$ is the Chevalley involution
for $x^{\vee}=x'$, and the hermitean Chevalley involution for $x^{\vee}=x^*$;
the subscript $0$ means the projection of $U({\cal G})$ on $U({\wp })$ which
is parallel to ${\cal G}_-U({\cal G})+U({\cal G}){\cal G}_+$. The given
definition is naturally extended for the case of the algebra $U'({\cal G})=
U({\cal G})\otimes_{U({\wp })}R({\wp }^*)$, where $R({\wp }^*)$ is the algebra
of the rational functions over ${\wp }^*$. It is very important to note that
the form $(x^{\vee} y)_0$ is degenerated on the left ideal $U'({\cal G}){\cal
G}_+$, and is not degenerated on the subalgebra $U({\cal G}_-)$; and hence is
not degenerated on the space $U'({\cal G})/U'({\cal G}){\cal G}_+$
which is a rational span of the corresponding Verma module. In fact, those
forms which are most relevant for describing the general solution to the Toda
type equations \cite{LS92}, are related with an appropriate holomorphic
extension
of the algebra $U({\cal G})$, namely $U_h({\cal G})=U({\cal G})\otimes_{U(
{\wp })}h({\wp }^*)$ where $h({\wp }^*)$ is the algebra of holomorphic
functions over ${\wp }^*$. Note that a special class of such extensions was
introduced in \cite{KK79}, see also \cite{K90}. In our case we deal with a
holomorphic extension of the algebra $U({\cal L})$, and the role
of the holomorphic and anitiholomorphic functions is played by the
functions $\Phi^{+}_m(z)$ and $\Phi^{-}_m(\bar{z})$, respectively, under
a relevant reality condition imposed, similarly to those in
\cite{RS94} for the Toda system, on the solution to equations (\ref{eqs}).

A basis in $U({\cal L})$ is constructed with the help of the monomials
$\hat{X}(\{ M\} )=X_{m_n}\cdots X_{m_1}$ in terms of the basis elements $X_{m}=
X_m(\Phi_m)$ of ${\cal L}$, see (\ref{prel}). In particular, let
$\hat{X}_{\pm}(\{M\} )$, $\hat{X}_{-}(\{ M\} )=\hat{X}_{+}^{\vee}(\{ M\} )$,
is such a basis in
$U({\cal G}_{\pm})$ with the weight $\mu (\{ M \} )$. Then the elements
$\hat{X}_+(\{ M\} )\hat{X}_-(\{ N\} )$ generate a basis of $U'({\cal L})$
over $R({\wp }^*)$, and this procedure gives for any weight $\mu\in
{\wp }^*$ a vector space $F_{\mu}({\cal L})$ of all formal series
$\sum_{\{ M,N\}}c_{\{ MN\}}\hat{X}_+(\{ M\} )\hat{X}_-(\{ N\} )$ with $c_{\{
MN\}}\in R({\wp }^*)$, where the sum runs over all monomials of the weight
$\mu =\mu (\{ M\} ) -\mu (\{ N\} )$. The subspaces $F_{\mu}({\cal L})$ are in
turn the subspaces of the algebra $F({\cal L})$ graded by the weights $\mu$.
To calculate explicitly the corresponding Shapovalov form or its hermitean
version, $(x^{\vee}y)$, one needs to transform the elements $\hat{X}_+(\{ M\}
)\hat{X}_-(\{ N\} )$ entering (\ref{tau}) to the series of the
monomials $\hat{X}_-\hat{X}_0
\hat{X}_+$, with $\hat{X}_0\in U({\wp})$, by using the commutation
relations (\ref{cr}) of the algebra ${\cal L}$. This is just the rule that
we have mentioned below formula (\ref{tau}). Now one can get convinced that
the matrix element (\ref{tau}) is a Shapovalov type form $(x^{\vee}y)_0$ for
some special two elements $x, y\in U_h({\cal G}_-)$ of the Lie algebra
${\cal L}$ in question.

\section{Acknowledgements}

Finishing up the paper, the author would like to thank E. Corrigan,
J.-L. Gervais, A. V. Razumov and A. M. Vershik for useful discussions. It is
a great pleasure for him to acknowledge the warm hospitality and creative
scientific atmosphere of the Laboratoire de Physique Th\'eorique de l'\'Ecole
Normale Sup\'erieure de Paris. This work was partially supported by the
Russian Fund for Fundamental Research and International Science Foundation.

\end{document}